\documentclass[11pt]{article}
\usepackage{latexsym}

\renewcommand{\@}[1]{\sqrt{#1}}

\def\be{\begin{eqnarray}}
\renewcommand{\le}[1]{\label{#1}\end{eqnarray}}
\def\ee{\end{eqnarray}}

\def\ffract#1#2{\raise .35 em\hbox{$\scriptstyle#1$}\kern-.25em/
\kern-.2em\lower .22 em \hbox{$\scriptstyle#2$}}

\begin{document}

\rightline{cond-mat/0805xx}
\begin{center}
\vspace{4.5cm} {\Large Using self-similarity and renormalization group to analyze time series}
\end{center}
\vspace{1cm}
\begin{center}
{\large Giovanni Arcioni \footnote{E-mail: {\tt
Giovanni.Arcioni@gmail.com, g.arcioni@noustat.it}} }\\
\vskip 1truecm
{\it Statistical and Mathematical Modeling\\
Noustat S.r.l,\\
 Via Pirelli 30, 20124 Milano, Italy }

\end{center}

\vspace{1.5cm}

\begin{center}
\large{Abstract}\\
\end{center}

An algorithm based on Renormalization Group (RG) to analyze 
time series forecasting was proposed in cond-mat/0110285. In this paper we explicitly code and  test it. We choose in particular some financial time series (stocks, indexes and commodities) with daily data and compute one step ahead forecasts.  We then construct some indicators to evaluate performances. The algorithm is supposed to prescribe the future development of the time series by using the self-similarity property intrinsically 
present in RG approach. This 
property could be potentially very attractive for the purpose of building winning trading systems. We discuss some relevant points along this direction. Although current performances have to be improved the algorithm seems quite reactive 
to various  combinations of input parameters and different past values sequences. This makes it a potentially good candidate to detect sharp market movements. We finally mention current drawbacks and sketch how to improve them. 

\newpage
\tableofcontents 
\section{Introduction}

Some years ago a novel approach based on Renormalization Group (RG) to analyze 
time series was proposed in \cite{YukalovMain}. An algorithm to make forecasts 
was outlined and some applications to financial markets were also examined in 
\cite{GluzmanYukalovA}, \cite{GluzmanYukalovB}.\\

The underlying idea of the algorithm is that 
the evolution of the market  can be formulated by means of a group property with respect to an ``effective" time, similarly to the evolution of a Hamiltonian with respect to its parameters in statistical renormalization  group. Sharp structural changes in 
the market, like booms and crashes, should be then  the equivalent of critical 
phenomena of a physical system (for reviews on renormalization group, phase 
transitions and critical phenomena see \cite{Zinn}, \cite{Parisi} , \cite{Goldendeld}).\\

In the present paper we explicitly test the algorithm proposed in \cite{YukalovMain} coding it in R language \cite{Rcran}.  We choose in particular some financial time series with daily data and compute one step ahead forecasts
(i.e. one day in this case). We then construct some indicators to make a performance evaluation of the algorithm. We also briefly outline some key points 
that one has to consider in order to use this or any other  algorithm to develop a (possibly) winning trading system (the author is currently working on building automated trading 
systems).\\

We refer for the general philosophy of the approach and details to \cite{YukalovMain}. 
However, there is one important point we would like to underline and seems to the author particularly interesting to 
investigate. The algorithm deeply relies on the concept of {\it self-similarity} which is 
intrinsically present in the RG philosophy. RG represents indeed a functional generalization of usual (i.e. power) self-similar transformations. This symmetry 
is {\it not} a symmetry of the physical system but a symmetry of the solution considered as  a function of the relevant physical variables (up to some specified boundary conditions). It is this symmetry which is supposed to provide the tool to detect the different ``regimes" 
of the time series under consideration, in this case bullish or bearish market movements, trends, 
crashes and so on.\\

As a consequence there is an important difference between traditional forecasting 
 models (like ARIMA,GARCH and similar) and the present one. In the former case one is 
instructed to compute the predicted value as a linear or more complicated  function of past times series values including in general some kind of noise. In the latter case, on the other hand, RG self-similarity is not an attempt to 
discover a relation between historical points, but tries to discover dynamic trends 
resulting in these points. This methodology  can be potentially very 
important in the case of financial time series.\\

The way in which self-similarity is used in this approach is also different from 
the way it is usually meant in connection with fractals: in that case, once a 
scaling relation is assumed, the function which describes the forecast is automatically 
computed; in the present case, on the other hand, one is referring to 
{\it group} self-similarity, therefore the equation which determines the function describing
the forecast (\ref{previsionefinale}) has to be solved as we are going to see in Section 2.\\

The paper is organized as follows: in Section 2 we review the algorithm proposed in 
\cite{YukalovMain}. In Section 3 we recall the application of the algorithm to time series. We simply present the final formulas which we have coded to test the 
algorithm. In Section 4 we report the results of the computer runnings. We first explain the kind of tests we have carried out and the types of inputs and indicators we have 
used. We explicitly show then the results of the tests. We end the Section with 
a brief excursus on trading systems to give a taste to the reader of the plethora of 
factors to be kept into account, beyond the algorithm itself, to build winning 
trading system. In Section 5 we discuss a statistical approach which we are currently testing and  should allow to weight in a consistent way the predictions of the algorithm. We end up in Section 6  with Outlooks and Conclusions. \\

\section{Review of the algorithm}

In this Section we review the main ideas and step underlying the algorithm and 
list the formulas which we have used to code it. We refer as 
said to \cite{YukalovMain} for a detailed and exhaustive discussion.\\

From a broad point of view the self-similar extrapolation used to make forecasts 
turns out to be at the end of the day an asymptotic expansion constructed out of past data. Let us 
review then briefly the various steps of the derivation, at a formal level first and 
then moving to the explicit example of time series.\\

Consider a function $f(x)$ and assume
\be
	f(x) \sim p_k (x) \hspace{0.3 cm} as \hspace{0.4cm} x \rightarrow 0
\ee
The  set of approximations $\{ p_k(x)\}$ is meaningful and convergent only 
asymptotically as $x \rightarrow 0$. It diverges for {\it finite} x. One therefore 
is not allowed to write
\be
	p_{k+1}(x) = f(p_k(x))
\ee
since the limit of $p_k(x)$ for $k \rightarrow \infty $ does not exists. Standard 
re-summation techniques, on the 
other hand, would require the knowledge of ten or more $p_k$. The self-similar approximation theory developed in \cite{YukalovMain} comes into help. The 
idea is to reorganize the sequence of approximations $p_k(x)$ in such a way to have 
convergence at finite x. This is done using renormalization.  
Formally one renormalizes with the help of a transformation $U$ the sequence of 
approximations $\{ p_k(x) \}$ according to
\be
U[p_k(x)] = \{ F_k(x,s) \}
\ee  
where $s$ are the so called control functions $s=s_k(x)$. $\{ F_k(x,s_k(x)) \}$ is 
now convergent. One defines then an expansion function $x=x(\phi,s)$ such that
\be
F_0(x,s)=\phi \Longrightarrow x=x(\phi,s)
\ee
It follows that
\be
y_k(\phi,s) = F_k(x(\phi,s),s) \Longrightarrow F_k(x,s)=y_k(F_0(x,s),s)
\ee
Therefore $\{ y_k(\phi,s) \}$ is in one to one correspondence with $\{ F_k(x,s) \}$. The limit $f^*(x)$ of 
$F_k(x,s_k(x))$ as $k \rightarrow \infty$ exists now by construction and this is the 
sought function $f(x)$. Due to this 1-1 correspondence this limit represents a fixed point for the $\{ y_k \}$ dynamics. In particular, near a fixed point, one can show that
\be
	y_{k+p} = y_k(y_p(\phi , s),s)
\ee
and this is nothing else than the self-similarity property. In continuous time one clearly has
\be
	y_k(\phi ,s) = y(t=k,s,\phi)
\ee
By means of self-similarity one can easily prove that 
\be
	\frac{\partial y(t,\phi, s)}{\partial t} = v(y(t,\phi,s))
\ee
This corresponds to the well-known $\beta$-function describing the RG flow. Integrating
\be
	\int_{y_k}^{y_{k+1}^*} \frac{dy}{v(y)} = \tau
\ee
where $y_k=y_k(\phi)$ and  $y_{k+1}^*=y_{k+1}^* (\phi,\tau)$. Using Euler discretization for the velocity one has
\be
v_k(\phi,s) = y_k(\phi,s)-y_{k-1}(\phi,s)
\ee
Consider now a concrete application of the previous steps. Let 
\be
f(x) \sim f_k(x) = \sum_{n=0}^k a_n x^{\alpha_n}
\ee
as $x\rightarrow 0.$ To get dimensionless units and scale invariance define then
\be 
\phi_k (x) = \frac{f_k(x)}{f_0(x)} = \sum_{n=0}^k b_n x^\beta_n
\ee
with $b_n=a_n/a_0$ and $\beta_n=\alpha_n - \alpha_0$. This expansion diverges for finite 
x. As explained before, apply then renormaliaztion. It is convenient to choose the control
functions using a power law transformation
\be
\label{PHI}
\Phi_k(x,s) = x^s \phi_k(x)
\ee
It follows that
\be
\Phi_k(x,s) = \sum_{n=0}^k b_n x^{s+\beta_n}
\ee
Now $\Phi_k(x,s)$ converges in $\mid x \mid < 1$ (as $s \rightarrow \infty$). Define then the expansion function $\phi$
\be
	\Phi_0(x,s) =\phi \Longrightarrow x=\phi^{1/s}
\ee
One easily finds 
\be
	y_k(\phi,s)=\Phi_k(x(\phi,s),s)= \sum_{n=0}^k b_n \phi^{1+\frac{\beta_n}{s}}
\ee
The ``effective time" $\tau_n$ to reach a fixed point after n steps is given by
\be
	\int_{n}^{n+\tau_n} dt = \int_{y_{n-1}}^{y_{n}^*} \hspace{0.2cm} 
	\frac{d\phi}{v_n(\phi,s)} 
\ee
One can show in a straightforward way that 
\be
\Phi_{n}^{*} = \left( \Phi_{n-1}^{-\beta_n / s}  -\frac{\beta_n}{s} 
 b_n \tau_n \right)^{-s/ \beta_n}
\ee
One then goes back to $\phi_k(x)$ using the inverse of (\ref{PHI}). Consider 
for instance the first order self similar approximation
\be
	\phi_{1}^{*} =\lim_{s \rightarrow \infty} x^{-s} \Phi_{1}^{*} (x,s)=
	\exp (b_1 \tau_1 x^\beta_1 )
\ee
Higher order approximations have been obtained in \cite{YukalovMain}
by means of a ``self-similar" bootstrap procedure. One gets at the end of the day
\be
	\phi_{k}^{*}(x) = \exp (c_1 x^{\nu_1} \exp (c_2 x^{\nu_2}...
	\exp c_k x^{\nu_k} ))... )
\ee
with 
\be
	c_n = \frac{a_n}{a_{n-1}} \tau_n , \hspace{0.4cm} 
	\nu_n = \alpha_n - \alpha_{n-1}  ,\hspace{0.4cm} 
	(n=1,2,...,k)
\ee
Notice that the effective time $\tau_n$ which appears in the previous 
formulas has not been defined yet. It was determined in \cite{YukalovMain} 
by using a time-distance cost functional F 
\be
\label{costfunction}
	F = \sum_n \left( (\tau_n - \frac{1}{n})^2 + (v_n \tau_n)^2 \right)
\ee
By minimizing it w.r.t. $\tau_n$ one has
\be
	\tau_n = \frac{1}{n(1+v_{n}^2)}
\ee
In a previous version of the algorithm \cite{GluzmanYukalovA}, the effective time was 
determined by a fixed point condition having the form of a minimal difference criterion. Such an equation, however, did {\it not} always have explicit solutions. We have 
checked this explicitly. We have also observed that the  effective time is in 
general {\it smaller} than the real time separation between time series data.\\

\section{Application to Time series}
We finally come to the application of the algorithm to time series. Following 
again the notation of \cite{YukalovMain} one arranges time series in backward 
recursion, i.e. one assumes $t_{n+1} < t_n$ with $(n=0,1,2,...,k)$ and the initial
time $t_0=0$. The past history is then given by
\be
\label{pasthistory}	
	D_k = \{ f_k,f_{k-1},...,f_0 \mid t_k < t_{k-1}<...<0 \}
\ee
In this case the sequence of approximations $f_k(t) $
\be
	f_k(t)= \sum_{n=0}^k a_n t^n \hspace{0.4cm} (a_0=f_0)
\ee
is such that it interpolates the given past history (\ref{pasthistory})
\be
	f_k(t_n)=f_n, \hspace{0.4cm} (n=0,1,2,...,k)
\ee
Using the same machinery discussed above one gets the final formula for the forecasts:
\be
\label{previsionefinale}
	f_{k}^* (t) = f_0 \exp \left( c_1 t \exp (c_2 t ... \exp(c_k t ))...\right)
\ee
with the controllers, the ``effective"  time and the ``velocity" respectively 
given by
\be
	c_n(t) = \frac{a_n}{a_{n-1}} \tau_n (t)
\ee
\be 
	\tau_n = \frac{1}{n[1+v_n^2(t)]}, \hspace{0.6cm}
	v_n (t) = \frac{a_n}{f_0} t^n, \hspace{0.4cm}
	(n=1,2,...,k)
\ee

This super-exponential function gives the forecast for the future time 
interval $[0,1]$ once the past history of $k$ values is given as input and it is supposed
 to incorporate the mixture of different 
tendencies and regimes of the time series. This is the formula that 
we have used in coding the algorithm in R language.\\

\section{Testing financial times series}
\subsection{Type of tests}

The results of the tests only provide of course  a limited 
case study, though we covered altogether almost five years of predictions.\\

First of all there is an almost infinite number of financial time series 
to be tested: one can study a stock, an index, a commodity or whatever for instance; in turn one can focus on stocks belonging to some specific sector; one can  choose assets according  to their correlation properties and so on. It depends of course on the targets one has. For instance, the choice could be motivated by some asset allocation procedure to 
construct a portfolio.\\

In addition, once the underlying has been chosen, one can study its time series depending on the time frames one wishes to focus on. Daily, monthly or annual  data can be 
some possible reasonable choices. One can also use
tick data and so on. Again, depending on the targets, one is 
supposed to make the most convenient choice among a huge number of possibilities.\\

In addition, even when one has chosen the set of time series to analyze, the tests have 
to cover as much as possible all the ``phase space" of available parameters and 
time intervals. A generic algorithm, for instance, can work well only in some specific 
conditions and/or periods. From this point of view we have tried to be quite exhaustive by exploring different choices of inputs.\\

Apart from considering a variety of financial time series, however, {\it the target
of these tests was to explore, as a first step, if the algorithm is reactive and how to different input
choices}. It turns out that it is and this is a positive sign.\\

We have therefore focused on daily data and computed one day ahead predictions (and also ``intraday", see 
below). We have studied  three  well known indexes: DAX, SP500 and VIX; one commodity, oil,  and finally two stocks: UBS and CreditSuisse. Before examining the results let us explain the parameters and indicators which we have used in the tests.\\

Examine first the inputs. First of all the number of past time series values. We have followed the choice made
 in \cite{GluzmanYukalovA} and \cite{GluzmanYukalovB} namely 3, 4, 5 and 6 points. This number of points ranging from 3 to 6 corresponds to the value 
of $k$ in the formula (\ref{previsionefinale}). Keeping in mind that each point basically adds an exponential in the formula
(\ref{previsionefinale}) going beyond six points does not seem to make much sense. As we observe below in Section 5 and already proposed in \cite{YukalovMain}, we have also 
considered the possibility to give as an input  not only a different number k of past values, but also different past-history sequences. So we have taken into account the most obvious sequence of past values at times $t=(-1,-2,-3,-4,-5,-6)$ (SequenceA), the sequences 
$t=(0,-1,-2,-3,-5,-8,-13)$(SequenceB-i.e. the Fibonacci series) and
$t=(0,-2,-4,-6,-8,-10,-12)$(SequenceC, i.e. evenly spaced values). Of course an infinite 
amount of choices is possible. Finally, for a one day prediction the time ahead is 1 (to be defined however as limit 
from the left according to the asymptotic expansion previously reviewed). We have also considered, 
however, the values 0.1, 0.3, 0.5, 0.7, i.e. which represents fractions of a day and therefore are to be considered 
``intraday" intervals. This is to test, once again, how reactive the algorithm is.\\

At the level of outputs we have constructed the following indicators: MAPE (i.e. the standard 
mean average percentage error) of the predicted values w.r.t to the real values. We have then computed 
the percentage number of cases in which APE (i.e. the absolute percentage error) between the predicted value and the real future value is less than the volatility 
of the future real value w.r.t the the previous past real value. We call this indicator
OKAPE, more precisely the ok is when APE is less than this 
volatility . We have computed than another indicator which tells the 
percentage number of cases in which 
the predicted values agree with up/down movement of the real value: for instance if the real future value is bigger than the past real one, is it true
that the predicted value as well is bigger than the past real one? We call this indicator \%TrendOK, since it tells us if the prediction ``gets the trend". This can 
be important for instance if one has in mind to construct trading strategies. It {\it can be the case} indeed that it is more fundamental to predict the correct trend rather than 
having a small MAPE. \\

In the following tables we have considered the inputs and the outputs discussed above doing at the same time an ``ensemble average" ( we return on this point in Section 5). 
More precisely: we have given as inputs  the number of points, the sequences and the times ahead discussed above all varying within their respective ranges and performed then the same forecast 
using all the possible combinations of the inputs' ranges, which amount to $4*3*5=60$, since we have used 4 training windows made of 3,4,5 or 6 points respectively, 3 sequences A,B,C of past points  and 5 time ahead intervals (0.1,0.3,0.5,0.7,1). So each time we make a prediction we have produced actually 60 possible outcomes. The best one each time 
could be determined having a suitable ``weight" acting on the space of predictions, as we will discuss in Section 5. As a first step, may be too crude but just to get an overall performance evaluation, we fix one of the three inputs and simply take the average of the other two. 
For instance, if the number of input points k is equal to 3, we take the 15 predictions
and compute their average.(15 since once k is fixed we do not have any more $4*3*5=60$ possibilities, but only $3*5=15$ remain). Same story with for the other parameters. The tables below are constructed according to this scheme. For instance, in 
the generic row SequenceA we will report the indicators that have been 
obtained once SequenceA is fixed, while k and the times ahead are free to change.
In this case each prediction will be the average of $3*5=15$ predictions and the
average indicators MAPE, OKAPE and \% TrendOK will be constructed out of these predictions; therefore they will be an average of an ensemble of averaged predictions. Let us move to the discussion of the specific cases.
\subsection{Results}
Consider the case study of the SP500 index \footnote{SP500 is an index containing the 
stocks of 500 large-cap corporations, most of which are American. }. We have done a test which covers the whole 2007. Here are the results.\\

\begin{tabular}{|r|r|r|r|}
\hline
         k &       MAPE & OKAPE &  \% TrendOK \\
\hline
         3 &      1.112 &       28.6 &       45.9 \\
         4 &      1.203 &       28.6 &       48.5 \\
         5 &       1.29 &       27.3 &       51.1 \\
         6 &      1.351 &         27 &       52.4 \\
\hline
 Sequence  &       MAPE & OKAPE &  \% TrendOK \\
\hline
         A &       1.07 &       33.3 &       51.2 \\
         B &      1.239 &       27.9 &       49.2 \\
         C &      1.408 &       22.3 &       48.1 \\
\hline
  Time.ahead &      MAPE & OKAPE &  \% TrendOK \\
\hline
       0.1 &      1.209 &       28.4 &       49.9 \\
       0.3 &      1.215 &       28.4 &       50.3 \\
       0.5 &      1.247 &       27.7 &       49.4 \\
       0.7 &      1.263 &       27.5 &       49.1 \\
         1 &      1.261 &       27.3 &       48.8 \\
\hline
\end{tabular}  
\\

MAPE is normally quite low (around 1 \%) but the percentage number of cases
in which it is smaller than the volatility (OKAPE) is quite low in all cases. 
Nevertheless, the percentage number of cases in which the correct trend is captured(
\% TrendOK) is above 50\% in some cases.\\

Consider now DAX index \footnote{DAX is a Blue Chip 
stock market index consisting of the 30 major German companies trading on the 
Frankfurt Stock Exchange.}  this time on a much shorter period, namely October 2007. The results are\\

\begin{tabular}{|r|r|r|r|}
\hline
		 k &       MAPE & OKAPE &  \% TrendOK \\
\hline
         3 &      0.819 &       21.8 &       37.6 \\
         4 &      0.933 &         24 &         43 \\
         5 &      1.061 &       24.5 &       43.6 \\
         6 &      1.159 &       20.9 &       37.9 \\
\hline
 Sequence&       MAPE & OKAPE &  \% TrendOK \\
\hline
         A &      0.937 &       19.5 &       30.7 \\
         B &      0.986 &       18.9 &       36.4 \\
         C &      1.055 &         30 &       54.5 \\
\hline
 Time.ahead&       MAPE & OKAPE &  \% TrendOK \\
\hline
       0.1 &      1.002 &       22.3 &       39.8 \\
       0.3 &      0.979 &       23.5 &       40.1 \\
       0.5 &      1.006 &       23.5 &       40.5 \\
       0.7 &      0.986 &       22.7 &       40.9 \\
         1 &      0.993 &         22 &       41.3 \\
\hline
\end{tabular}  
\\

MAPE is normally below one percent, but the other two indicators do not display 
positive results. Consider now VIX index \footnote{VIX is the ticker symbol for the Chicago Board 
Options Exchange Volatility Index, a popular measure of the implied volatility of SP500 index options. It can be thought as a sort of thermometer of the market's expectation of volatility.}. The test has been 
performed on the whole 2007. The results are\\

\begin{tabular}{|r|r|r|r|}
\hline
     k &       MAPE & OKAPE &  \% TrendOK \\
\hline
         3 &      9.614 &       33.5 &       50.2 \\
         4 &     10.141 &       34.1 &       52.3 \\
         5 &     10.073 &       34.5 &       53.7 \\
         6 &     10.057 &       35.9 &         54 \\
\hline
 Sequence&       MAPE & OKAPE &  \% TrendOK \\
\hline
         A &      8.882 &         37 &       52.5 \\
         B &      9.867 &       34.7 &       51.5 \\
         C &     11.165 &       31.8 &       53.7 \\
\hline
 Time.ahead&       MAPE & OKAPE &  \% TrendOK \\
       0.1 &       9.54 &       35.9 &       53.8 \\
       0.3 &      9.595 &       35.4 &       53.3 \\
       0.5 &      9.896 &       34.8 &       52.7 \\
       0.7 &     10.351 &       33.7 &       51.9 \\
         1 &     10.475 &       32.6 &       51.2 \\
\hline
\end{tabular}  
\\

Both MAPE and OKAPE are not particularly good. Note, however, that the \% TrendOK indicator
is always above 50\%. In one case even 54\%. This is a clear example in which 
despite the fact MAPE is quite big, the prediction gets the right trend more then half of the 
times.\\

Consider now a commodity, namely oil. The tests have been carried out for the 
whole 2007. Here are the results\\

\begin{tabular}{|r|r|r|r|}
\hline
     k &       MAPE & OKAPE &  \%TrendOK \\
\hline     
         3 &      2.457 &       34.4 &       51.1 \\
         4 &      2.677 &       30.6 &       50.1 \\
         5 &      2.978 &         29 &       50.7 \\
         6 &      3.278 &       26.3 &       48.8 \\
\hline
 Sequence&       MAPE & OKAPE &  \%TrendOK \\
\hline
         A &      2.511 &       33.1 &       51.1 \\
         B &      2.844 &       31.3 &       51.2 \\
         C &      3.187 &       25.7 &       48.2 \\
\hline
 Time.ahead&       MAPE & OKAPE &  \%TrendOK \\
\hline
       0.1 &      2.854 &       29.8 &         50 \\
       0.3 &      2.856 &       29.6 &         50 \\
       0.5 &       2.86 &       29.9 &       50.3 \\
       0.7 &      2.842 &       30.3 &       50.5 \\
         1 &      2.826 &       30.6 &       50.1 \\
\hline
\end{tabular}  
\\

Again MAPE and OK APE are not quite satisfactory, but the trend indicator seems
to give promising results. We finally move than to stocks: we consider UBS and Credit Suisse for the period 
from 14 Jan to 29 Feb 2008, in which both stocks experienced quite dramatic draw-downs. A challenging test for the algorithm then. In the case of UBS one has\\

\begin{tabular}{|r|r|r|r|}
\hline
     k &       MAPE & OKAPE &  \% TrendOK \\
\hline
         3 &      4.706 &       30.5 &       48.8 \\
         4 &       5.98 &       25.3 &       43.6 \\
         5 &      7.385 &       18.1 &       41.7 \\
         6 &      8.138 &       18.3 &       43.8 \\
\hline
 Sequence&       MAPE & OKAPE &  \% TrendOK \\
\hline
         A &      5.248 &       23.4 &       40.7 \\
         B &      7.585 &       15.7 &       36.7 \\
         C &      6.824 &         30 &         56 \\
\hline
 Time.ahead&       MAPE & OKAPE &  \% TrendOK \\
\hline
       0.1 &      6.761 &       22.1 &       43.1 \\
       0.3 &      6.735 &       22.4 &       44.1 \\
       0.5 &       6.59 &       22.6 &       44.8 \\
       0.7 &      6.346 &         24 &         45 \\
         1 &      6.329 &       24.1 &       45.5 \\
\hline
\end{tabular}  
\\ 

while for the case of Credit Suisse\\

\begin{tabular}{|r|r|r|r|}
\hline
     k &       MAPE & OKAPE &  \%TrendOK \\
\hline
         3 &      3.647 &       35.2 &       51.7 \\
         4 &      4.684 &       28.7 &         56 \\
         5 &      5.566 &       24.4 &       52.1 \\
         6 &      6.409 &       15.7 &       46.7 \\
\hline
 Sequence&       MAPE & OKAPE &  \%TrendOK \\
\hline
         A &      4.047 &       27.6 &       50.3 \\
         B &      5.805 &       21.7 &       44.7 \\
         C &      5.377 &       28.8 &       59.9 \\
\hline
 Time.ahead&       MAPE & OKAPE &  \%TrendOK \\
\hline
       0.1 &       5.37 &       25.2 &       52.3 \\
       0.3 &      5.287 &       25.5 &       52.5 \\
       0.5 &      5.214 &       25.3 &       51.3 \\
       0.7 &       4.92 &       25.8 &       50.8 \\
         1 &       4.59 &       28.3 &       51.3 \\
\hline
\end{tabular}  
\\

Notice MAPE and OKEAPE are more or less similar; actually the two time series 
are quite correlated, so this fact does not sound  so unreasonable.  In the case of CreditSuisse, 
however, the algorithm is much more successful  in getting the trend, up to 
59.9\% in one case!.\\

Altogether the results  are not yet completely satisfactory. In the 
next Section we will sketch the steps to improve them. The algorithm, 
however, is very ``reactive" according to the different types of inputs. The back tests above, in addition, {\it have been carried out using  a ``rolling forward" training modality}, i.e. the past history of k-points is rolled forward as soon as one moves ahead to 
compute the predictions. {\it At each prediction, therefore, the algorithm makes the prediction
according to a rolling forward training window and the mixture of exponential reacts 
each time to capture the time series tendencies}.\\

We have also used a standard optimization procedure to see if there are ``magic combintations" or some preferred parameters. The results are shown in the table below where the entries express the percentage number of times in which the corresponding row parameter best performed.\\

\begin{tabular}{|c|c|c|c|c|c|c|}
\hline 
 Parameter &      SP500 &        DAX &        VIX &        OIL &        UBS &        CS \\
 \hline
       k=3 &       36.2 &       45.5 &       35.6 &       36.1 &         40 &       45.5 \\
       k=4 &       21.9 &       22.7 &         16 &       21.3 &       31.4 &       27.3 \\
       k=5 &         21 &       22.7 &         20 &       19.7 &       11.4 &       15.2 \\
       k=6 &         21 &        9.1 &       28.4 &       22.9 &       17.1 &       12.1 \\
\hline
      SeqA &       35.7 &       18.2 &       33.2 &       33.7 &       28.6 &       33.3 \\
      SeqB &       17.6 &       40.9 &       16.4 &       20.1 &       22.9 &       30.3 \\
      SeqC &       46.7 &       40.9 &       50.4 &       46.2 &       48.6 &       36.4 \\
\hline
      Time.ahead.0.1. &       56.7 &       27.3 &         64 &       55.4 &       51.4 &       30.3 \\
       Time.ahead.0.3. &        5.7 &        4.5 &          2 &        3.2 &        5.7 &          3 \\
       Time.ahead.0.5. &        4.8 &        9.1 &        5.6 &        5.6 &        5.7 &          3 \\
       Time.ahead.0.7. &        7.6 &        4.5 &        6.4 &        7.2 &       14.3 &      21.2 \\
       Time.ahead.1    &       25.2 &       54.5 &         22 &       28.5 &       22.9 &       42.4 \\
\hline
\end{tabular}  
\\

One can notice that the best results would have been obtained using in all cases
k equal to three as far as the number of training points concerns. This seems to 
be in agreement with the general philosophy of RG flows: collective coherent behavior
should require only a few renormalization steps to get convergence much like the case 
of phase transitions. The most intuitive sequenceA i.e. past values at times $t=0,-1,-2...,-6$ is not the best one.\\
 
\subsection{Intermezzo: trading systems and trading performance reports}

In the next Section we discuss possible improvements of the performances of the algorithm. Before moving to these aspects, we briefly recall some essential points concerning 
trading systems. The algorithm has been applied to financial time series but of course, 
apart from the outputs and relative error between real future data and predicted values, the 
essential point in this context is what to do with the forecasts and which strategy 
one is supposed to implement.\\

A considerable number of factors has indeed to be taken into account and according to them 
one can establish if the prediction is ``good" or ``bad".  Many of these points are actually not commonly discussed in the mathematical 
finance literature (see for instance \cite{Hull}, \cite{Joshi}), but they do play a fundamental role.  We mention just a few of them to give a taste to  the reader.\\

Commissions costs and splippages for instance. It can be that an algorithm detects the right patterns and is able to predict them in the future. However, the potential gain can be eaten by commissions and splippage costs. Consider then time frames: for example intraday or daily. One algorithm could give good daily predictions and be suitable for daily trading then but not for intraday trading. Another point: although one algorithm could give not quite satisfactory results , still, however, in suitable diversified portfolio, the net profit could be positive. And so on.\\

There are then certain indicators which appear in almost all trading strategy performance reports which are of fundamental importance: total net profit, number of winning/losing trades, largest winning/losing trade, average winning/losing trade, max drawdown, max consecutive winners/losers and so on. Again many factors can influence these final indicators and the forecasting algorithm is only one of them.\\

This is of course an extremely! brief and not exhausting discussion of trading systems 
but suffices to show that many factors apart from the predictive algorithm have to be taken into account. Therefore, even if the algorithm performed extremely well, it would not be obvious that one would be able to build a winning strategy. On the other hand, even if the results of the algorithm do not seem good enough, may be, if they combined with 
other tools, they can lead to a winning trading system. For instance: in all cases in which the indicator \%TrendsOK constructed before  is above 50\%, although MAPE is high, it could be a good starting point to construct a winning strategy. For more details on trading systems we refer the interested reader to \cite{Natenberg}, 
\cite{Stridsman}, \cite{katz}, \cite{hill}.\\

\section{Improving performances: a statistical mechanics of possible scenarios}

The results we have shown in the previous Section have to be improved. One natural way 
would be to define an averaged forecast weighted by some suitable probabilities. In a way, it amounts to do  statistical mechanics with a partition function and a probability 
measure.\\

In a sense the scheme used to produce the tables shown before already goes along this direction: we do not compute a single prediction each time, but an ensemble of predictions according to the possible 
different combinations of the inputs, which turn out to be 60 in our case. We have made then a performance evaluation analysis simply taking the averages: 
once we fix one of the three inputs (k or Sequence or Times ahead) we consider all the forecasts obtained by letting the other two change and take an average. This is a too crude way of 
weighting however and it is not able to pick out the best combinations.\\

Indeed let us reconsider the previous tests. This time, however, we consider the $4*3*5=60$ possible combinations of the inputs without fixing any of then and averaging over the other two as we did before. Take DAX as an example. We show here only some of the most significant combinations\\ 

\begin{tabular}{|r|r|r|r|r|r|}
\hline
     k &  Sequence &   Time.ahead &       MAPE &  OKAPE &  \%trendOK \\
\hline
          3 &          3 &        0.1 &      0.701 &       27.3 &       54.5 \\
         3 &          3 &        0.3 &      0.674 &       36.4 &       54.5 \\
         3 &          3 &        0.5 &      0.786 &       36.4 &       59.1 \\
         3 &          3 &        0.7 &      0.761 &       31.8 &       54.5 \\
         3 &          3 &          1 &      0.686 &       31.8 &       54.5 \\
         4 &          3 &        0.1 &      0.963 &       36.4 &       63.6 \\
         4 &          3 &        0.3 &      0.914 &       36.4 &       63.6 \\
         4 &          3 &        0.5 &      0.834 &       36.4 &       63.6 \\
         4 &          3 &        0.7 &      0.777 &       36.4 &       68.2 \\
         4 &          3 &          1 &      0.873 &       36.4 &       68.2 \\
            5 &          3 &        0.1 &      1.099 &       31.8 &       54.5 \\
         5 &          3 &        0.3 &      1.052 &       31.8 &       54.5 \\
         5 &          3 &        0.5 &      1.229 &       31.8 &       54.5 \\
         5 &          3 &        0.7 &      1.187 &       31.8 &       59.1 \\
         5 &          3 &          1 &      1.179 &       27.3 &       54.5 \\
         
\hline
\end{tabular}  

As one can notice for $k=4$ there are some combinations which capture the trend more than 
60 \% of the times and this is a very positive outcome. The same is true for the other 
underlings. Consider for instance UBS. Here is a list of good combinations which 
correctly get the trends

\begin{tabular}{|r|r|r|r|r|r|}
\hline
		 k &	Sequence &    Time.ahead & MAPE & OKAPE & \% TrendOK \\
\hline 
        4 &          3 &        0.1 &      5.482 &       34.3 &       62.9 \\
         4 &          3 &        0.3 &      5.831 &       34.3 &       62.9 \\
         4 &          3 &        0.5 &       5.88 &       34.3 &       57.1 \\
         4 &          3 &        0.7 &      6.176 &       37.1 &       57.1 \\
         5 &          3 &        0.3 &      7.586 &       28.6 &       62.9 \\
         5 &          3 &        0.5 &      7.071 &       28.6 &       62.9 \\
         5 &          3 &        0.7 &      6.958 &       25.7 &       57.1 \\
         5 &          3 &          1 &      7.615 &       22.9 &       57.1 \\
\hline
\end{tabular}

A similar story holds for the other assets.  The natural question is then how 
to pick up the best forecast out the different ones obtained with different inputs. In other words one should find a way to compute weights which are able to pick up the best predictions. This is currently under investigation. The general philosophy, however, was already outlined in Section 4 of \cite{YukalovMain}.\\

In a nutshell, the past history has to take into account not only the number of past 
points k but also different time scales. To fix ideas define the past data base as  ensemble of possible scenarios 
\be
	D_k(j) =\left( f_{k}^{(j)},f_{k-1}^{(j)},...,f_0 \mid 
         t_{k}^{(j)} < t_{k-1}^{(j)}< ... <0  \right)
\ee
where k stands as usual for the number of past points, while j specifies the chosen time sequence for the past (The analogue of the choice of SequencesA,B,C in the tests we have done). One however has to define a probability weight $p_k(j,t)$, which will be able to select the best forecast out of the combinations of the (k, j, t)
values chosen. Notice that this is just a more compact form to express the same 
choice of inputs we have used in our tests.\\

One proposal could be the following
\be
	p_k (j,t) = \frac{1}{Z_k(t)} \exp \left( - \Delta S_k (j,t)\right) \hspace{0.6cm}
	Z_k (t) = \sum_j \exp (- \Delta S_k (j,t) )
\ee
where
\be
	\Delta S_k (j,t) = S_k(j,t) - S_1 (j,t), \hspace{0.6cm}
	S_k (j,t) = \ln \mid \delta f_{k}^* (j,t) \mid
\ee
with $S_k(j,t)$ representing a dynamical entropy. Define then a bunch of multipliers $m_k (j,t)$ as 
follows
\be
	\Delta S_k (j,t) = \ln \mid m_k (j,t) \mid
\ee
and an average multiplier $\bar{m_k(t)}$
\be
	\frac{1}{\bar{m_k(t)}} = \sum_j \frac{1}{\mid m_k(j,t) \mid}
\ee
One can easily show that
\be
p_k (j,t) = \mid \frac{\bar{m_k(t)}}{m_k(j,t)} \mid 
\ee
The most probable scenario is given by
\be
	\max_{j} p(j,t) 
\ee
The forecast will be then a weighted sum of the different forecasts
\be
	<f(t)> = \sum_j \sum_k p_k(j,t) f_{k}^{*}(j,t)
\ee
with t as usual in the interval $[0,1]$.

\section{Outlooks and Conclusions}

We have tested the algorithm on different financial time series. The results can 
certainly be improved and in the previous Section we have outlined a possible path, which
should give the right way to weight different forecasts obtained using different inputs. \\

There are however other points which are currently under investigation  and which 
could improve the algorithm; we list them briefly:\\

$\bullet$ In order to determine the effective time $\tau$ one has to minimize the cost function (\ref{costfunction}). The choice made in \cite{YukalovMain} seems reasonable, but other choices 
are of course possible as well and this could improve the performances of the algorithm.\\

$\bullet$ One can of course give as an input to the algorithm not the time series itself, but some 
sequence obtained by transforming the original time series. The log of the original time 
series or the differenced time series for instance are standard transformations which 
could be worth checking.\\

$\bullet$ Filter the original time series before making the forecasts. 
One could use wavelets and apply multiresolution analysis to decompose according to different time scales (The R package waveslim provides all the 
necessary functions to investigate this point).\\

$\bullet$  Considering external variables could be also a very promising path to explore. The external 
force, however, as recalled in \cite{GluzmanYukalovA}, has to be not too strong, otherwise 
it could break self-similarity properties. In the opinion of the author it is however quite 
tough to measure in a quantitative way the strength of the external forces so it is 
definitively worth trying to add external or even categorical variables to the model.\\

$\bullet$ The number of points given as input in our tests ranges between 3 and 6. Of course
one could use more, but then the final formula (\ref{previsionefinale}) would contain too many exponentials. Of course one may object that a training window with a maximum number of 6 points is not enough to learn the past history: think for instance about neural networks which need quite massive training sets to learn and become then predictive. It depends however on the targets one has. From the author's point of view, the main interest to investigate this algorithm, apart from the conceptual one of extending applications of the RG, is to find out if it is quite reactive enough to detect trends, possibly without too much delay. From this perspective a maximum number of  six training points perhaps is not too low.\\

All these issues are currently under investigation.

\section{Acknowledgments }

The author would like to thank Maurizio Sanarico for reading the paper and important remarks, 
Marco Pomponi for really many detailed and extremely useful explanations on trading systems and their software implementation. 
He also acknowledges Sorin Solomon, Alessandro Cappellini and 
Vyacheslav Yukalov  for discussions and correspondence. .  

\end{document}